# HVSRweb: An Open-Source, Web-Based Application for Horizontal-to-Vertical Spectral Ratio Processing

Joseph P. Vantassel, M.Sc. S.M. ASCE,[1] Brady R. Cox, Ph.D., P.E., M. ASCE,[2] and
Dana M. Brannon S.M. ASCE[3]

[1]The University of Texas at Austin, Dept. of Civil, Arch. and Envir. Eng., 301 E. Dean Keeton St. Austin, TX 78712, USA E-mail: jvantassel@utexas.edu
[2]The University of Texas at Austin, Dept. of Civil, Arch. and Envir. Eng., 301 E. Dean Keeton St. Austin, TX 78712, USA E-mail: brcox@utexas.edu
[3]The University of Texas at Austin, Dept. of Civil, Arch. and Envir. Eng., 301 E. Dean Keeton St. Austin, TX 78712, USA. E-mail: dana.brannon@utmail.utexas.edu

## ABSTRACT

The horizontal-to-vertical spectral ratio (HVSR) method has become an increasingly popular tool for developing a quick and reliable estimate of a site's fundamental natural frequency ($f_0$). This paper presents HVSRweb, an open-source, web-based application for performing HVSR calculations in a convenient, reliable, and statistically-consistent manner. HVSRweb allows the user to upload three-component ambient noise records and perform the HVSR calculation in the cloud, with no installation required. The HVSR calculation can be performed using a single combined horizontal component (e.g., geometric-mean, squared-average) or multiple rotated horizontal components to investigate azimuthal variability. HVSRweb can reject spurious time windows using an automated frequency-domain window-rejection algorithm, removing the need for subjective and time-consuming user interaction. It also facilitates the use of lognormal statistics for $f_0$, allowing for a consistent statistical framework capable of quantifying measurement uncertainty in terms of either frequency or its reciprocal, period. In addition, HVSRweb presents the opportunity to rapidly incorporate new developments in processing and quantification of uncertainty as they emerge, and encourages standardization of HVSR processing across users while avoiding the challenges associated with traditional approaches that require the user to perform regular updates to keep pace with new developments.

## INTRODUCTION

The horizontal-to-vertical spectral ratio (HVSR) of ambient noise is a fast and non-intrusive method for estimating a site's fundamental natural frequency ($f_{0,site}$). Many studies (Lachet & Bard, 1994; Lermo & Chávez-García, 1993) have shown that when a strong impedance contrast is present, the HVSR peak frequency ($f_{0,HVSR}$) closely corresponds to $f_{0,site}$. First proposed by Nogoshi and Igarashi (1971) and later popularized by Nakamura (1989), the HVSR method has gained widespread use due to its simplicity in field acquisition and data processing. The field acquisition involves placing a three-component sensor (typically a broadband seismometer) on the ground surface and leaving it to record ambient vibrations for some period of time. The amount of time a sensor is left to record is often on the order of tens of minutes to hours. The data processing begins with dividing ambient noise records into smaller records, called windows. Windows are typically between 30 seconds and a few minutes in length, with longer time windows being recommended for locations with lower $f_{0,site}$ (SESAME, 2004). In practice the window length is usually selected first and in accordance with the anticipated $f_{0,site}$. With the window length selected, the total record length required in the field is selected to ensure a statistically significant number of windows, typically 30 or more. Once the time record is broken into smaller windows, the two horizontal components from each window are often combined into a single horizontal component either before (e.g., azimuthal rotation) or after (e.g., geometric-mean, squared-average) being transformed to the frequency domain. Finally, the division of the horizontal and vertical Fourier amplitude spectrum results in the HVSR curve. If there is a clear peak (or peaks) in the HVSR curve, the lowest frequency peak provides an estimate of $f_{0,site}$. By repeating this procedure for all of the time windows, it is possible to develop statistics for $f_{0,HVSR}$ (and therefore $f_{0,site}$). For the remainder of the work, we will refer to $f_{0,HVSR}$ and $f_{0,site}$ as $f_0$.

This paper presents HVSRweb, an open-source, web-based application for performing HVSR processing of ambient noise. The back-end of HVSRweb is *hvsrpy,* which is an open-source Python package for HVSR processing (Vantassel, 2020). *hvsrpy* was developed to implement new HVSR processing techniques proposed by Cox et al. (2020) and Cheng et al. (2020). We strongly encourage the reader to review these aforementioned works for background information relating to HVSRweb's novel functionalities, the presentation of which is the primary focus of this paper. These novel functionalities include a frequency-domain window-rejection algorithm (FWA), which allows for the automatic rejection of contaminated time windows in the frequency-domain, the ability to incorporate azimuthal variability into the statistical representation of $f_0$, and the use of lognormal statistics, which allow for consistent representation of uncertainty, regardless of whether $f_0$ or its reciprocal, fundamental site period ($T_0$), is the desired parameter of interest. HVSRweb is hosted in collaboration with the DesignSafe-CI (Rathje et al., 2017) and can be accessed at https://hvsrweb.designsafe-ci.org/. We strongly encourage the reader to interact with the application as they read this work.

**MOTIVATION FOR A WEB APPLICATION**

Before detailing the functionality of HVSRweb, the benefits derived from a web-application over the two most common alternatives, processing scripts and executable programs, deserve a brief discussion. These benefits include: ease of access, scientific repeatability, and reduction in development latency. First, concerning ease of access, web applications do not require the download or configuration of software and are not dependent on the user's operating system. In contrast, processing scripts and executable programs will always depend on one or both. Furthermore, web applications need not be confined to a single machine, allowing access from any device provided an internet connection is available. Second, in regards to scientific repeatability, a web-based application is rather similar to a standard executable program, in that the implementation details have been encapsulated from the user. With processing scripts, however, the front-end (user interface) is entwined with the back-end (implementation details), which allows a user to edit the program (wittingly or otherwise). A web-application avoids this by separating the user interface and the implementation details. Finally, web-applications offer an unmatched ability to reduce the latency in the distribution of future developments. In traditional executable programs, or processing scripts, the user must check when new versions become available and devote time to the installation/setup and, as is most often the case, resolve upgrade conflicts and/or consider how to manage existing projects which are dependent on previous versions. All of these issues are immediately resolved with a web-application, as the trouble of updating the software and providing access to earlier versions is the sole responsibility of the application developers and not the user. Once the developer releases a new feature, it is immediately available to all users without the need for the user to download, install, or update anything. For these reasons, we considered a web-application the most appropriate way to develop a universal platform for HVSR processing that is easy-to-access and easy-to-use.

**APPLICATION DEVELOPMENT**

HVSRweb was developed by incorporating aspects from a number of open-source projects. Namely, *hvsrpy* and Plotly's Dash framework. *hvsrpy* is an open-source Python package for HVSR processing (Vantassel, 2020) which provides the back-end to HVSRweb. As the remainder of the paper focuses primarily on the functionality of HVSRweb, and therefore a subset of the functionality available in *hvsrpy,* we will not discuss the details of it here. Instead, we refer the interested reader to the *hvsrpy* page on the Python Package Index (https://pypi.org/project/hvsrpy/) for useful links and benchmark test cases. Of particular note are the many comparisons between *hvsrpy* and the popular open-source software Geopsy (Wathelet et al., 2020). These examples illustrate *hvsrpy*'s commitment to exactly reproduce the results of Geopsy, wherever the processing parameters and functionality of *hvsrpy* and Geopsy overlap. This was done to allow users to check *hvsrpy* and to encourage standardization in HVSR processing. The front-end of HVSRweb was developed using Plotly's Dash framework (Plotly, 2017). Dash is an open-source library that allows web applications to be written in pure Python; no HTML, CSS, or JavaScript required. Dash itself is built on top of other open-source projects, which include Flask, a popular Python-based micro web framework (Ronacher, 2015), Plotly.js, a JavaScript library of interactive plotting components (Plotly, 2012), and React.js, a JavaScript library of reactive application components (Facebook Inc., 2013). Leveraging Dash for the user-interface and *hvsrpy* for the back-end allows HVSRweb to contain all the aspects users expect from a web application (e.g., buttons, drop-down menus) but with only a fraction of the traditional development resources required. This allowed HVSRweb to be built by two of the authors (DMB & JPV) with limited web-application experience in only a few months of part-time development.

**FUNCTIONALITY**

HVSRweb was designed to make HVSR processing as customizable as possible, thereby allowing the user to decide on the most appropriate settings for their particular project. Of course, as a result, we risk overcomplicating the application and overwhelming the novice user. The remaining sections of this work seek to mitigate this risk by discussing the available options, providing guidance on best practices, and highlighting those settings with the greatest impact on the result to ensure they receive special consideration. The key settings are discussed below in three sections that correspond with three tabs used in HVSRweb's user interface: Time, Frequency, and HVSR.

**TIME-DOMAIN SETTINGS**

The settings in the Time tab include: window length, cosine taper width, and time-domain filter. Recall the first step of HVSR processing of ambient noise is the division of a long time record into individual time windows. SESAME (2004) recommends using time windows that are at a minimum ten times $1/f_0$ (i.e., ten times $T_0$). However, we strongly recommend using longer time windows whenever possible, and windows that are preferably no shorter than 30 seconds. To ensure a sufficient number of windows for developing meaningful statistics on $f_0$, we recommend a minimum recording time of ambient noise include 40 to 60 time window lengths. Currently, HVSRweb only accepts three-component time records saved in the miniSEED format, however, the authors intend to add accommodations for other formats as they are identified by users. In order to help the reader visualize the effects of various processing settings discussed in this work, a 60 minute time record (i.e., UT.STN11.A2_C150.miniseed) from a dataset published by Cox and Vantassel (2018) is used throughout this work. The file can be loaded using HVSRweb's Demo button, located in the upper-left corner. Note that by providing this data file we enable the interested reader to reproduce the figures presented in this work, as well as explore the effects of other settings discussed but not presented visually.

Returning to the main discussion, panels (a), (b), and (c) of Figure 1 illustrate HVSR processing using window lengths of 30, 60, and 120 seconds, respectively, with all other settings following the predefined defaults in HVSRweb, except the FWA option in the HVSR tab (discussed later), which has not been used for this example. Interestingly, while the individual HVSR curves from each time window vary widely between the three window lengths, the median HVSR curve and its 68% confidence interval (CI) do not vary significantly. However, as discussed below, since the statistics of $f_0$ are dependent on the individual time windows (Cox et al., 2020) (and not the median curve), an appropriate window length must be selected carefully. After splitting the time record into windows, the ends of the windows are cosine tapered using a Tukey window (commonly referred to as a cosine-tapered window) to prevent spectral leakage artifacts from appearing in the frequency domain. The width of the cosine taper can vary between 0 (rectangular window) and 1 (Hann window), with 0.1 (i.e., 5% off either end of the window) being recommended. While the decision to use a taper generally has an effect on the HVSR curve, the exact value is typically not of significance provided a reasonable value is selected (e.g., between 0.05 and 0.2). The final time domain settings are related to a Butterworth bandpass filter. The filter can be tuned in terms of its upper and lower cut-off frequencies and order. Butterworth filtering is generally not necessary for records of good quality; however, the option is provided for the user's convenience, though it is deselected by default. Should the user wish to use the Butterworth filter, the minimum and maximum frequencies should be selected carefully to ensure that low or high frequency HVSR peaks are not being excluded. For the example presented in Figure 1, a reasonable frequency bandwidth would be between 0.1 and 30 Hz (i.e., just outside the HVSR curve plot limits). A filter order of 5 is recommended.

**FREQUENCY-DOMAIN SETTINGS**

There are two settings in the Frequency tab: the Konno and Ohmachi (1998) smoothing coefficient (*b*) and the desired frequency resampling. As the name implies, *b* controls the relative smoothness of the frequency domain curves and, therefore, that of the HVSR curves. To illustrate, three different values for *b* (20, 40, and 60) were used to develop the HVSR curves shown in Figure 2a, b, and c, respectively. Again, all other settings follow the predefined defaults except the FWA, which has not been used. Figure 2 illustrates that a smaller *b* results in smoother curves, with the median curve clearly being affected. For most applications, a value of 40 or 50 is recommended (Cox et al., 2020). When specifying the frequency resampling parameters, care must be taken to ensure that the minimum and maximum frequencies are low and high enough, respectively, to encompass all HVSR peaks. The number of frequency points between the minimum and maximum has a minor effect on the HVSR curve, provided that a sufficient number of points are considered. At least 128 logarithmically-spaced points is recommended.

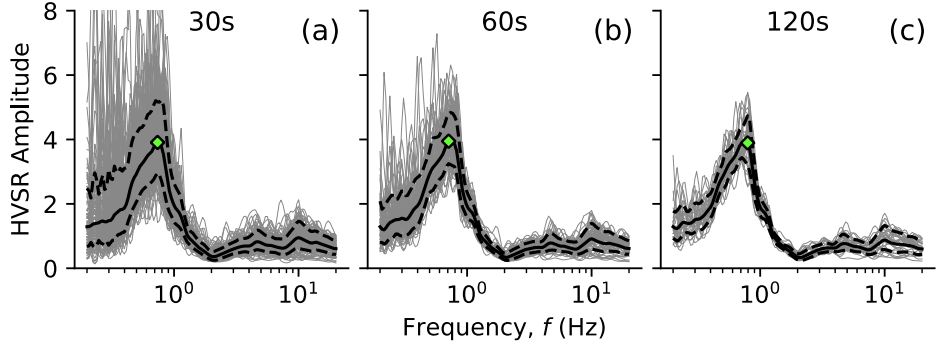

**Figure 1.** *Effect of window length on HVSR* processing when using: (a) 30-, (b) 60-, and (c) 120-second long time windows. The gray lines are the HVSR curves from each time window, the solid black line is the lognormal median curve, the green diamond is the peak of the median curve ($f_{0,mc}$), and the dashed black lines represent the 68% confidence interval of the HVSR curve.

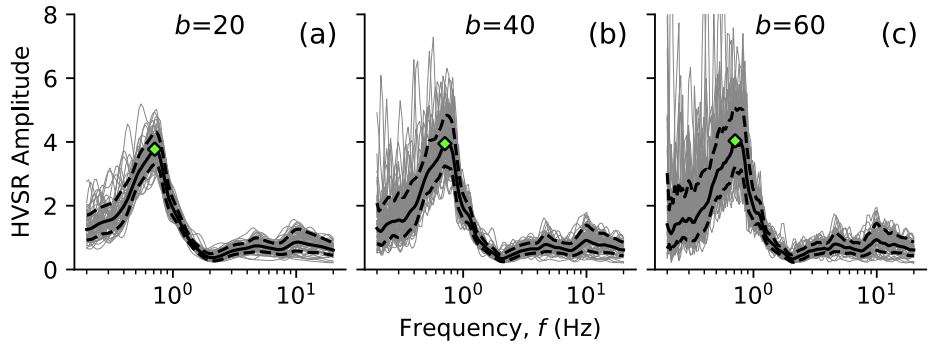

**Figure 2.** *Effect of the Konno and Ohmachi smoothing coefficient (b) on HVSR* processing considering values of: (a) 20, (b) 40, and (c) 60. Refer to the caption of Figure 1 for a full description of the various features shown.

**HVSR-PROCESSING SETTINGS**

The web application includes four settings in the HVSR tab: definition of the horizontal component, distribution of $f_0$, distribution of median curve, and FWA settings. As mentioned previously, the two horizontal components must be combined. To do this, a number of methods have been proposed, including the geometric-mean, the square-average, rotation to a single azimuth, and rotation through multiple azimuths. All of these methods have been included in HVSRweb, and the first three (the fourth will be discussed separately) are illustrated in Figure 3. All other settings follow the predefined defaults, except the FWA, which has not been used in this example, but is discussed further below. Note that the three HVSR median curves are distinct in terms of their shape, amplitude, and frequency peak. To provide a clear suggestion to the reader, we will echo the recommendations of Cox et al. (2020) and Cheng et al. (2020), namely, that if a single HVSR curve is desired, then the lognormal median curve ($LM_{curve}$) calculated using the geometric-mean should be used. However, if a statistical representation of $f_0$ is desired, then the multiple-azimuth approach (option four, discussed below) should be used. For the distributions of $f_0$ and the median curve, we recommend lognormal, following Cox et al. (2020). The lognormal distribution of $f_0$ is of particular import, as it allows the statistics determined in terms of frequency to be exactly related in terms of period, which is not possible using a normal distribution. The final HVSR settings pertain to the FWA, the details of which are presented in Cox et al. (2020). However, a short discussion will be made here. First, the user defines a parameter *n*, which controls the permissiveness of the FWA. A value of *n* equal to 2.0 standard deviations is recommended by Cox et al. (2020), but *n* could be increased by the user to 2.5 or 3.0 if they feel too many good windows are being rejected. Second, all individual windows with an $f_0$ beyond *n* standard deviations from $LM_{f0}$ are rejected. $LM_{f0}$ is then recalculated using only the accepted $f_0$ values and the rejection process is repeated until reaching the convergence criteria. In addition to *n*, *hvsrpy* and, therefore, HVSRweb also requires the user to set a maximum number of iterations. In our experience, the FWA generally requires fewer than 15 iterations to converge, and so the default number of allowable iterations has been set conservatively to 50. The user should ensure the iterations to completion are less than the allowable.

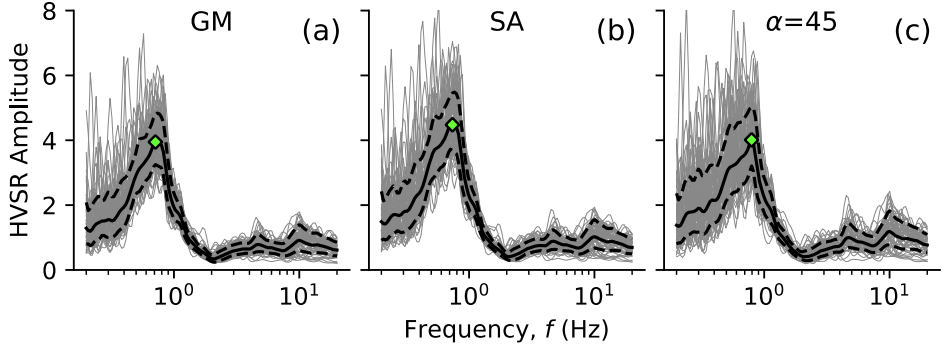

**Figure 3.** *Effect of combining the horizontal components* **using: (a) the geometric-mean (GM), (b) the squared-average (SA), and (c) an azimuth 45 degrees anti-clockwise from north ($\alpha = 45$). Refer to the caption of Figure 1 for a full description of the various features shown.**

**EXAMPLES**

To fully demonstrate HVSRweb, the example ambient noise record is first processed using the geometric-mean following Cox et al. (2020), and second by rotating the horizontal components through multiple azimuths at a given azimuthal interval (AI) following Cheng et al. (2020). The geometric-mean approach is summarized in Figure 4 and Table 1, while the multiple-azimuths approach is summarized in Figure 5 and Table 2. Note that both examples use the default values provided by HVSRweb. Figure 4b, shows the individual HVSR curves and the lognormal median HVSR curve ($LM_{curve}$) before application of the automated FWA. Careful inspection reveals a number of low frequency time-window peaks ($f_{0,i}$) which are biasing the lognormal median of the $f_0$ values ($LM_{f0}$) to a lower value that is inconsistent with the peak of the lognormal median curve ($f_{0,mc}$). However, in Figure 4d, the FWA is shown to correctly remove these contaminated windows, bringing $f_{0,mc}$ and $LM_{f0}$ into reasonable agreement. To visualize the time-domain recording and illustrate which time windows were rejected, the entire noise record for the three components (i.e., north-south (NS), east-west (EW), and vertical (VT)) are shown in Figure 4a, 4c, and 4e, respectively. Note that the rejected time-windows are not obvious outliers in terms of their time-domain response, giving further justification for the use of a frequency-domain rejection algorithm.

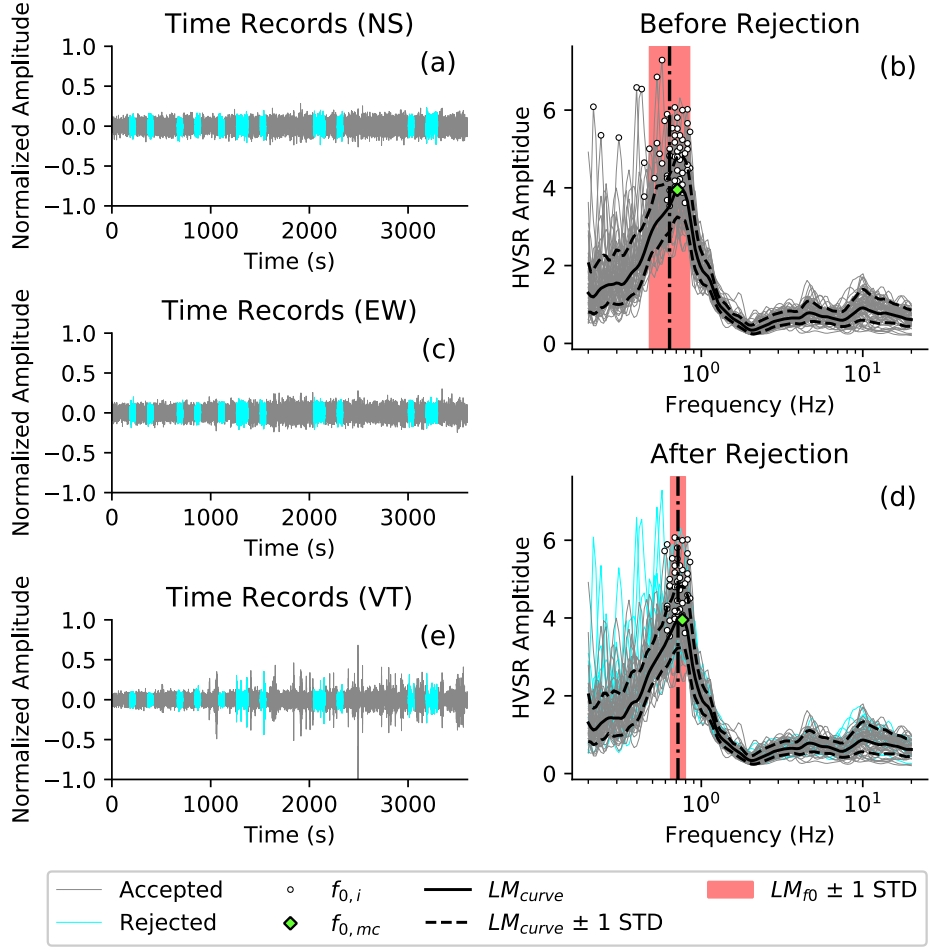

**Figure 4. Example ambient noise record processed using HVSRweb considering the *geometric-mean*.** Time domain recordings of ambient noise for the three orthogonal components [i.e., (a) north-south (NS), (c) east-west (EW), and (e) vertical (VT)] with the windows automatically rejected by the frequency-domain window-rejection algorithm (FWA) shown in cyan. Individual HVSR curves from each time window (b) before application of the rejection algorithm, and (d) after its application. Also shown are the lognormal median curve ($LM_{curve}$) and uncertainty ($LM_{curve} \pm 1\ STD$), peaks of each accepted time window ($f_{0,i}$), the peak of the median curve ($f_{0,mc}$), and the uncertainty of the fundamental site frequency ($LM_{f0} \pm 1\ STD$).

**Table 1. HVSR statistics after FWA rejection considering the *geometric-mean horizontal component*.**

|  | *Lognormal Median (LM)* | *Lognormal Standard Deviation ($\sigma_{ln}$)* |
|---|---|---|
| *Fundamental Site Frequency, $f_{0,GM}$* | 0.72 Hz | 0.10 |
| *Fundamental Site Period, $T_{0,GM}$* | 1.39 s | 0.10 |

**Table 2. HVSR statistics after FWA rejection considering *multiple horizontal azimuthal components*.**

|  | *Lognormal Median (LM)* | *Lognormal Standard Deviation ($\sigma_{ln}$)* |
|---|---|---|
| *Fundamental Site Frequency, $f_{0,AZ}$* | 0.68 Hz | 0.18 |
| *Fundamental Site Period, $T_{0,AZ}$* | 1.48 s | 0.18 |

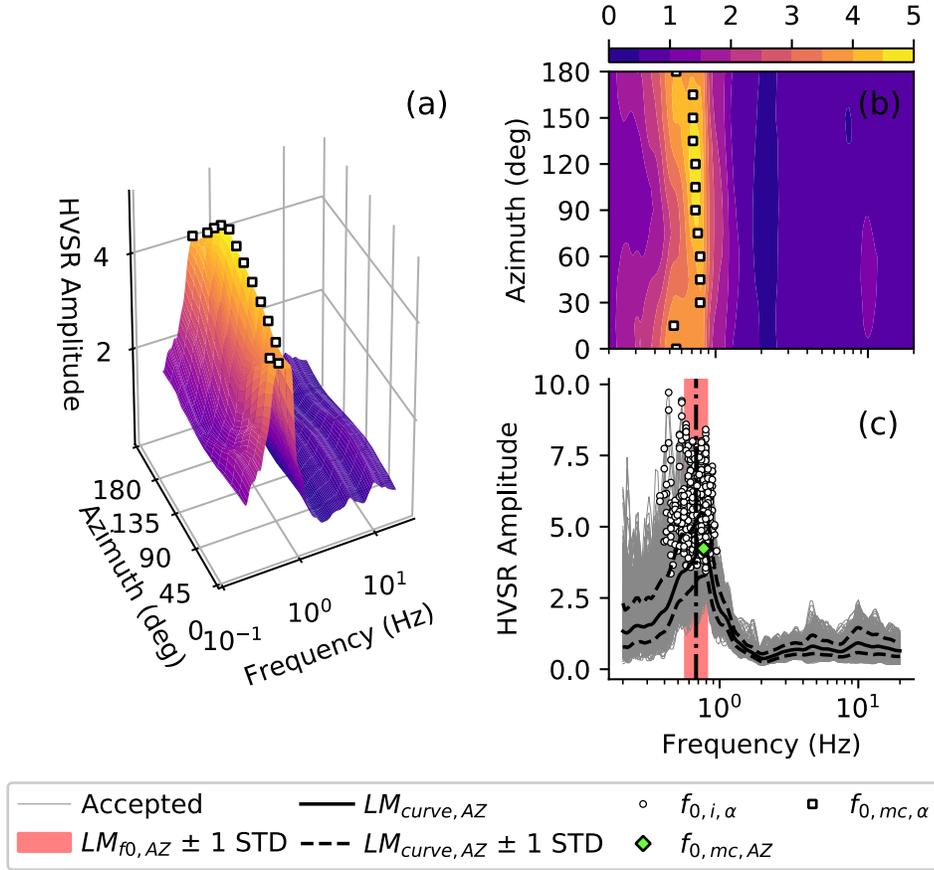

**Figure 5. Example ambient noise record processed using HVSRweb considering *multiple azimuths*. Panel (a) shows the median HVSR curve at each azimuth in frequency-azimuth-amplitude space. Panel (b) shows a two-dimensional projection of panel (a) with the peak of the lognormal median curve at each azimuth ($f_{0,mc,\alpha}$) clearly indicated. Panel (c) shows the accepted HVSR curves from each azimuth and every time window after applying the frequency-domain window-rejection algorithm (FWA). Also shown is the azimuthally-averaged (AZ) lognormal median curve ($LM_{curve,AZ}$) and its uncertainty ($LM_{curve,AZ} \pm 1\ STD$), peaks of each accepted window for every azimuth ($f_{0,i,\alpha}$), the peak of the AZ lognormal median curve ($f_{0,mc,AZ}$), and the uncertainty of the AZ fundamental site frequency ($LM_{f0,AZ} \pm 1\ STD$).**

Figure 5 illustrates the multiple-azimuth approach to HVSR processing, which allows for azimuthal variability to be quantified in the statistical representation of $f_0$. In particular, Figure 5a shows a three-dimensional (i.e., frequency-azimuth-amplitude) plot of the lognormal median HVSR curve at different azimuths (α). Figure 5b shows a projection of Figure 5a onto the frequency-azimuth plane to better illustrate the change of $f_{0,mc}$ as a function of azimuth (i.e., $f_{0,mc,\alpha}$). Figure 5c shows the individual HVSR curves from every accepted time window along all azimuths. The peak of each window ($f_{0,i,\alpha}$) is also shown, along with the azimuthally-averaged (AZ) lognormal median curve ($LM_{curve,AZ}$), the peak of the AZ lognormal median curve ($f_{0,mc,AZ}$), and the uncertainty on the AZ fundamental site frequency ($LM_{f0,AZ} \pm 1\,STD$). A qualitative comparison between Figures 4 and 5, and a quantitative comparison between Tables 1 and 2, show that the AZ procedure for estimating $f_0$ has larger uncertainty [i.e., an increase of the lognormal standard deviation ($\sigma_{ln}$) by 0.08] and is shown to better capture the changes of $f_0$ across all azimuths (refer to Figure 5b). In contrast, their median curves (i.e., $LM_{curve}$ and $LM_{curve,AZ}$) are quite similar in terms of amplitude, shape, and peak frequency with $f_{0,mc}$ and $f_{0,mc,AZ}$ both occurring at approximately 0.76 Hz. However, their median curves are not similar in terms of their uncertainty with the AZ procedure showing increased uncertainty over the GM procedure. HVSRweb produces a short summary of the most pertinent results in the Results tab. This output includes the time window length, the number of iterations to convergence of the FWA, and the total number of windows rejected. In addition, HVSRweb also displays the statistical information from the time window peaks before and after rejection in a similar format to Tables 1 and 2. And, while not a statistic, $f_{0,mc}$ is also provided. In addition, HVSRweb offers several downloadable outputs including: publication-ready figures, such as those shown in Figures 4 and 5, a highly-detailed summary file that includes the statistics on $f_0$ and the median curve, and an alternate summary file which follows the Geopsy-style output format to ease the transition of those who have already developed HVSR workflows using Geopsy.

## CONCLUSION

This paper presents HVSRweb (https://hvsrweb.designsafe-ci.org/)*,* an interactive web-based application for HVSR processing. HVSRweb allows the user to process HVSR data in the cloud without needing to download, install, or update any software. The only requirements for using HVSRweb are a modern internet browser and an internet connection. HVSRweb allows flexibility in how the HVSR processing is to be performed, leaving it to the user to decide which settings are most appropriate for their project. However, as the number of choices can be overwhelming for the uninitiated, HVSRweb provides reasonable default values for all settings. This paper seeks to provide context to those recommended values, as well as guidance on when they should be changed. Several examples of HVSR processing using HVSRweb are presented to familiarize the reader with the available functionality. By making HVSRweb easy-to-access and easy-to-use, the authors' hope to encourage a more universal standard of practice and provide a platform for seamless introduction of future developments in HVSR processing and quantification of uncertainty.

## ACKNOWLEDGEMENTS

The first author would like to thank Dr. Krishna Kumar for his recommendation to investigate Plotly and his generosity for hosting HVSRweb during its early development. The authors would like to thank Dr. Ellen Rathje and the DesignSafe team, Salvador Tijerina in particular, for their generosity and assistance in permanently hosting HVSRweb on the DesignSafe-CI (https://hvsrweb.designsafe-ci.org/). To encourage reproducibility, all previous versions of HVSRweb and instructions for accessing and running them locally are provided on the project's GitHub (https://github.com/jpvantassel/hvsrweb).